\documentclass[twocolumn,prd,preprintnumbers]{revtex4-1}
\usepackage{amsmath,amssymb,graphicx,booktabs,bm,psfrag,color,slashed,euscript}

\newcommand{\F}{{\EuScript F}}
\newcommand{\G}{{\EuScript G}}
\newcommand{\J}{{\EuScript J}}

\begin{document}

\title{\boldmath Evolution of the $B\hspace{0.3mm}$-Meson Light-Cone Distribution Amplitude in Laplace Space}

\preprint{MITP/20-031}
\preprint{June 9, 2020}

\author{Anne Mareike Galda$^a$}
\author{Matthias Neubert$^{a,b}$}

\affiliation{${}^a$PRISMA$^+$\! Cluster of Excellence {\rm \&} Mainz Institute for Theoretical Physics (MITP), Johannes Gutenberg University, 55099 Mainz, Germany\\
${}^b$Department of Physics {\em\&} LEPP, Cornell University, Ithaca, NY 14853, U.S.A.}

\begin{abstract}
The $B$-meson light-cone distribution amplitude is a central quantity governing non-perturbative hadronic dynamics in exclusive $B$ decays. We show that the information needed to describe such processes at leading power in $\Lambda_{\rm QCD}/m_b$ is most directly contained in its Laplace transform $\tilde\phi_+(\eta)$. We derive the renormalization-group (RG) equation satisfied by this function and present its exact solution. We express the RG-improved QCD factorization theorem for the decay $B^-\to\gamma\ell^-\bar\nu$ in terms of $\tilde\phi_+(\eta)$ and show that it is explicitly independent of the factorization scale. We propose an unbiased parameterization of $\tilde\phi_+(\eta)$ in terms of a small set of uncorrelated hadronic parameters.
\end{abstract}

\maketitle

\section{Introduction} 

Light-cone distribution amplitudes (LCDAs) describe the inner structure of hadrons as probed in hard exclusive QCD processes. They are non-perturbative quantities of fundamental importance for the theory and phenomenology of the strong interactions \cite{Lepage:1979zb,Lepage:1980fj,Efremov:1978rn,Efremov:1979qk,Chernyak:1983ej}. The LCDAs of heavy hadrons appear, e.g., in calculations of heavy-hadron pair production at $e^+ e^-$ colliders \cite{Grozin:1996pq} and in the study of symmetry relations between the form factors describing transitions between heavy and light mesons \cite{Beneke:2000wa}. The leading-order LCDA of the $B$ meson, $\phi_+^B(\omega)$, plays a particularly prominent role in the QCD factorization approach to exclusive non-leptonic decays such as $B\to M_1 M_2$ and $B\to M\gamma$, where $M_i$ denote light mesons ($m_{M_i}\ll m_B$) \cite{Beneke:1999br,Beneke:2000ry,Beneke:2001at}. The corresponding decay amplitudes can be written as the sum of two terms, one in which the relevant hadronic information is encoded in experimentally accessible form factors $F_i^{B\to M}(q^2)$, and a ``hard-scattering'' contribution governed by the $B$-meson LCDA. The relative size of the two terms depends on a hadronic parameter $\lambda_B$ defined in terms of a weighted integral over the LCDA.

The radiative decay $B^-\to\gamma\,\ell^-\bar\nu$ offers a particularly clean probe of the LCDA, because in this case the form factor term is absent \cite{Lunghi:2002ju,Bosch:2003fc} (see also \cite{Korchemsky:1999qb,DescotesGenon:2002mw}). At leading order in an expansion in powers of $\Lambda_{\rm QCD}/m_b$ the decay amplitude for this process can be written as 
\begin{equation}\label{fact}
   {\cal M}(E_\gamma) \propto m_B f_B \int_0^\infty\!\frac{d\omega}{\omega}\,
    T(m_b,E_\gamma,\omega,\mu)\,\phi_+^B(\omega,\mu) \,,
\end{equation}
where $E_\gamma<m_B/2$ is the energy of the photon as measured in the rest frame of the $B$ meson. The hard-scattering kernel $T$ can be factorized further as
\begin{equation}
   T(m_b,E_\gamma,\omega,\mu) = H(m_b,E_\gamma,\mu)\,J(-2E_\gamma\omega,\mu) \,.
\end{equation}
The process can be treated in QCD factorization as long as $2E_\gamma\sim m_b$, where $m_b\simeq 4.8$\,GeV denotes the pole mass of the bottom quark. The decay amplitude is sensitive to three different energy (or distance) scales: a ``hard'' scale set by the $b$-quark mass, the scale of non-perturbative QCD dynamics ($\omega\sim\Lambda_{\rm QCD}$), and an intermediate scale of order $\sqrt{m_b\Lambda_{\rm QCD}}$. The hard function $H$, the jet function $J$ and the LCDA $\phi_+^B$ contain the contributions from these three hierarchical scales in factorized form. The first two of these functions can be calculated in QCD perturbation theory, while the LCDA is a genuinely non-perturbative object, which can be defined in terms of the hadronic matrix element \cite{Grozin:1996pq,Lange:2003ff} 
\begin{equation}
   \langle 0|\,\bar q\,\delta(\omega-in\cdot\!\overleftarrow{D})\,
    \rlap{\hspace{0.2mm}/}{n}\gamma_5 h_v|\bar B(v)\rangle
   = i F(\mu)\,\phi_+^B(\omega,\mu) 
\end{equation}
in heavy-quark effective theory. Here $h_v$ denotes the heavy-quark spinor field, $v^\mu$ is the 4-velocity of the $B$ meson and $n^\mu$ is a lightlike vector satisfying $v\cdot n=1$. The hadronic parameter $F(\mu)$ is related to the decay constant $f_B$ of the $B$ meson via $\sqrt{m_B}\,f_B=K_F(m_b,\mu)\,F(\mu)$, up to power corrections of order $\Lambda_{\rm QCD}/m_b$. $K_F$ is a perturbative matching coefficient, which is part of the hard function $H$. The factorization scale $\mu$ in (\ref{fact}) is arbitrary and in principle can be chosen at will, but there is no single scale choice for which the hard-scattering kernel $T$ is free of large logarithms. The resummation of these logarithms can be accomplished by solving the renormalization-group (RG) evolution equations for $H$, $J$ and $\phi_+^B$. For $\phi_+^B$ and beyond next-to-leading order (NLO) in perturbation theory, this is however a formidable task.

When the hard-scattering kernel $T$ is calculated in fixed-order perturbation theory, all that is probed of the LCDA are the logarithmic moments \cite{Beneke:2011nf}
\begin{equation}\label{moments}
\begin{aligned}
   \frac{1}{\lambda_B(\mu)}
   &= \int_0^\infty\!\frac{d\omega}{\omega}\,\phi_+^B(\omega,\mu) \,, \\
   \sigma_n(\mu)
   &= \lambda_B(\mu) \int_0^\infty\!\frac{d\omega}{\omega}\,
    \ln^n\!\Big(\frac{\bar\omega}{\omega}\Big)\,\phi_+^B(\omega,\mu) \,,
\end{aligned}
\end{equation}
where $\bar\omega$ serves as a fixed reference scale. If a specific model for the LCDA is assumed, these moments can be expressed in terms of a few model parameters. For example, the simple exponential model $\phi_+^B(\omega)=(\omega/\omega_0^2)\,e^{-\omega/\omega_0}$ \cite{Grozin:1996pq} yields $\lambda_B=\omega_0$, $\sigma_1=\gamma_E-\ln\frac{\omega_0}{\bar\omega}$ and $\sigma_2=\frac{\pi^2}{6}+\big(\gamma_E-\ln\frac{\omega_0}{\bar\omega}\big)^2$ etc. This imposes an important limitation: Because the correlations between the different moments are highly model dependent, it is difficult to assess the uncertainties in these relations. 

In contrast to the LCDAs of light mesons, not much is known on general grounds about the properties of the $B$-meson LCDA. In particular, the function $\phi_+^B(\omega,\mu)$ does not approach a simple asymptotic form in the formal limit $\mu\to\infty$, and the integral over this function is divergent \cite{Grozin:1996pq}. It has, however, been shown that for sufficiently large values of $\mu$ the LCDA scales like $\omega$ for $\omega\to 0$ and falls off slower than $1/\omega$ for $\omega\to\infty$ \cite{Lange:2003ff}. Several models for $\phi_+^B(\omega,\mu)$ have been proposed in the literature. Some of them are based on QCD sum-rule estimates \cite{Grozin:1996pq,Ball:2003fq,Braun:2003wx}, while others are inspired by an {\em ad hoc\/} modeling of the LCDA in momentum space \cite{Lee:2005gza} or in the so-called ``dual space'' \cite{Bell:2013tfa,Feldmann:2014ika,Beneke:2018wjp}, where its one-loop evolution equation takes on a simpler form. Most of these models rest on unjustified assumptions, which imply important biases and lead to uncontrolled systematic uncertainties: 
i) The LCDA is often assumed to be positive definite, even though it is an {\em amplitude\/} that does not admit a probabilistic interpretation. In fact, it has been argued that $\phi_+^B(\omega,\mu)$ changes sign for some value of $\omega\gg\Lambda_{\rm QCD}$ \cite{Braun:2003wx,Lee:2005gza}, which implies that the moments $\sigma_n$ can have either sign, even if $n$ is an even integer. 
ii) Many models assume that at a low renormalization scale $\mu_s$ the LCDA exhibits an exponential fall-off for large $\omega\gg\Lambda_{\rm QCD}$, even though this is in conflict with RG evolution. At best, this assumption could be true at one particular value of $\mu_s$, but RG evolution to a scale $\mu>\mu_s$ inevitably leads to a fall-off slower than $1/\omega$ \cite{Lange:2003ff}. 
iii) Any given model for $\phi_+^B(\omega,\mu)$ necessarily implies strong correlations between the moments $\sigma_n$, for which there is no reason {\em a priori}.

In this work we show that the information that can be probed in hard exclusive processes is entirely and most directly described by the {\em Laplace transform\/} of the LCDA, defined as
\begin{equation}\label{Laplace}
   \tilde\phi_+(\eta,\mu) = \int_0^\infty\!\frac{d\omega}{\omega}\,\phi_+^B(\omega,\mu)
    \left( \frac{\omega}{\bar\omega} \right)^{-\eta} .
\end{equation}
In fixed-order calculations one probes the behavior of the Laplace transform near the origin, since
\begin{equation}\label{origins}
   \frac{1}{\lambda_B(\mu)} = \tilde\phi_+(0,\mu) \,, \qquad
   \frac{\sigma_n(\mu)}{\lambda_B(\mu)} = \tilde\phi_+^{(n)}(0,\mu) \,,
\end{equation}
where the superscript denotes the $n^{\rm th}$ derivative with respect to the first argument. More generally, solving the RG equation for the jet function one finds that at leading order $J(-p^2,\mu)\propto\left(p^2/\mu_j^2\right)^{a_\Gamma(\mu_j,\mu)}$ \cite{Bosch:2003fc}, where $\mu_j\sim\sqrt{m_b\Lambda_{\rm QCD}}$ is a suitable matching scale and the exponent $a_\Gamma(\mu_j,\mu)$ will be defined below. The decay amplitude in (\ref{fact}) is thus determined directly by the Laplace transform $\tilde\phi_+\!\left(-a_\Gamma(\mu_j,\mu),\mu\right)$ evaluated at a point away from the origin. We show that also beyond the leading order all hadronic information needed to calculate the decay rate is encoded in this function. Moreover, we derive an explicit expression for the factorized decay amplitude in (\ref{fact}), in which the factorization scale $\mu$ drops out explicitly and large logarithms are resummed to all orders of perturbation theory.

\section{RG Evolution in Laplace Space} 

We write the RG evolution equation for the $B$-meson LCDA in the general form \cite{Lange:2003ff}
\begin{equation}\label{RGEphiB}
   \frac{d\phi_+^B(\omega,\mu) }{d\ln\mu}
   = - \int_0^\infty\!d\omega'\,\gamma_+(\omega,\omega';\mu)\,\phi_+^B(\omega',\mu) \,,
\end{equation}
with the anomalous dimension
\begin{equation}\label{gammaplus}
\begin{aligned}
   \gamma_+(\omega,\omega';\mu) 
   &= \left[ \Gamma_c(\alpha_s)\,\ln\frac{\mu}{\omega} + \gamma(\alpha_s) \right] 
    \delta(\omega-\omega') \\[1mm]
   &\quad\mbox{}- \Gamma_c(\alpha_s)\,\omega\,\Gamma(\omega,\omega') 
    - \hat\gamma_+(\omega,\omega';\alpha_s) \,.
\end{aligned}
\end{equation}
The coefficient $\Gamma_c$ of the logarithm is the lightlike cusp anomalous dimension in the fundamental representation of $SU(N_c)$ \cite{Korchemskaya:1992je}. The same quantity appears in the coefficient of the symmetric plus distribution
\begin{equation}
   \Gamma(\omega,\omega') = \left[ \frac{\theta(\omega-\omega')}{\omega(\omega-\omega')}
    + \frac{\theta(\omega'-\omega)}{\omega'(\omega'-\omega)} \right]_+ ,
\end{equation}
which is defined such that, when $\Gamma(\omega,\omega')$ is integrated with a function $f(\omega')$, one must replace $f(\omega')\to f(\omega')-f(\omega)$ under the integral. The function $\hat\gamma_+$ starts at two-loop order (see below). 

On dimensional grounds, the terms shown in the second line of (\ref{gammaplus}) can be written as $1/\omega$ times a function of the ratio $x=\omega'/\omega$. It will be useful to define the dimensionless functions
\begin{align}\label{calGdef}
   \F(\eta) 
   &= \int_0^\infty\!dx\,\Gamma(1,x)\,x^\eta = - \big[ H(\eta) + H(-\eta) \big] \,, \notag \\
   \G(\eta;\alpha_s)
   &= \int_0^\infty\!dx\,\hat\gamma_+(1,x;\alpha_s)\,x^\eta \,,
\end{align}
where $H(\eta)=\psi(1+\eta)+\gamma_E$. Based on (\ref{RGEphiB}) we find that the Laplace transform of the LCDA obeys the non-linear, partial differential equation
\begin{equation}\label{wonderful}
\begin{aligned}
   & \left( \frac{d}{d\ln\mu} 
    + \Gamma_c(\alpha_s)\,\frac{\partial}{\partial\eta} \right) \tilde\phi_+(\eta,\mu) \\[1mm]
   &= \bigg[ \Gamma_c(\alpha_s) \left( \!\ln\frac{\bar\omega}{\mu} + \F(\eta)\! \right)
    - \gamma(\alpha_s) + \G(\eta,\alpha_s) \bigg]\,\tilde\phi_+(\eta,\mu) \,,
\end{aligned}
\end{equation}
which is analogous to an equation for the soft-quark soft function in radiative Higgs-boson decay derived in \cite{Liu:2020eqe}. While this equation appears rather intimidating at first sight, its exact solution can be found by noting that any function of the combination $\eta+a_\Gamma(\mu_0,\mu)$, with
\begin{equation}\label{aGamma}
   a_\Gamma(\mu_0,\mu) 
   = - \int\limits_{\alpha_s(\mu_0)}^{\alpha_s(\mu)}\!
    d\alpha\,\frac{\Gamma_c(\alpha)}{\beta(\alpha)} 
   \approx \frac{2C_F}{\beta_0}\,\ln\frac{\alpha_s(\mu)}{\alpha_s(\mu_0)}
\end{equation}
and for some fixed scale $\mu_0$, is a solution of the homogeneous equation with the right-hand side of (\ref{wonderful}) set to zero. Here $\beta(\alpha_s)$ is the QCD $\beta$-function, $\beta_0=11-\frac23\hspace{0.3mm}n_f$ (with $n_f=4$ light quark flavors in our case) and $C_F=\frac43$. Let $\mu_s$ denote the matching scale at which the boundary condition $\tilde\phi_+(\eta,\mu_s)$ is defined. The ansatz
\begin{equation}\label{solu}
\begin{aligned}
   &\tilde\phi_+(\eta,\mu)
    = N(\mu_s,\mu)\,\tilde\phi_+\big(\eta+a_\Gamma(\mu_s,\mu),\mu_s\big) \\[1mm]
   &\quad\times
    \exp\Bigg[\, \int\limits_{\alpha_s(\mu_s)}^{\alpha_s(\mu)}\!\frac{d\alpha}{\beta(\alpha)}\,
    \bigg[ \Gamma_c(\alpha)\,\F(\eta+a_\Gamma(\mu_\alpha,\mu)\big) \\[-3mm]
   &\hspace{3.5cm}\mbox{}+ \G\big(\eta+a_\Gamma(\mu_\alpha,\mu),\alpha\big) \bigg] \Bigg] \,,
\end{aligned}
\end{equation}
where $\mu_\alpha$ is defined such that $\alpha_s(\mu_\alpha)\equiv\alpha$, then provides the solution to (\ref{wonderful}) with the correct boundary condition if we require that the normalization $N(\mu_s,\mu)$ satisfies the differential equation
\begin{equation}
   \frac{dN(\mu_s,\mu)}{d\ln\mu} 
   = \bigg[ \Gamma_c(\alpha_s)\,\ln\frac{\bar\omega}{\mu} - \gamma(\alpha_s) \bigg]\,
    N(\mu_s,\mu) \,,
\end{equation}
with the initial condition $N(\mu_s,\mu_s)=1$. This yields
\begin{equation}
   N(\mu_s,\mu) 
   = \left( \frac{\bar\omega}{\mu_s} \right)^{-a_\Gamma(\mu_s,\mu)} 
    e^{\,S(\mu_s,\mu) + a_\gamma(\mu_s,\mu)} \,,
\end{equation}
where the quantity $a_\gamma$ is defined in analogy with (\ref{aGamma}) and
\begin{equation}\label{Sdef}
   S(\mu_s,\mu) 
   = - \int\limits_{\alpha_s(\mu_s)}^{\alpha_s(\mu)}\!d\alpha\,
    \frac{\Gamma_c(\alpha)}{\beta(\alpha)}
    \int\limits_{\alpha_s(\mu_s)}^\alpha\!\frac{d\alpha'}{\beta(\alpha')} \,.
\end{equation}
The integral over the function $\F$ in (\ref{solu}) can be evaluated by changing variables from $\alpha$ to $a_\Gamma(\mu_\alpha,\mu)$. This leads to the exact solution
\begin{align}\label{greatsolu}
   \tilde\phi_+(\eta,\mu) 
   &= N(\mu_s,\mu)\, 
    \frac{\Gamma\big(1+\eta+a_\Gamma(\mu_s,\mu)\big)\,\Gamma(1-\eta)}%
         {\Gamma\big(1-\eta-a_\Gamma(\mu_s,\mu)\big)\,\Gamma(1+\eta)} \notag \\
   &\times \exp\Bigg[\,\int\limits_{\alpha_s(\mu_s)}^{\alpha_s(\mu)}\!\frac{d\alpha}{\beta(\alpha)}\,
    \G\big(\eta+a_\Gamma(\mu_\alpha,\mu),\alpha\big) \Bigg] \notag \\[2mm]
   &\times e^{2\gamma_E a_\Gamma(\mu_s,\mu)}\,\tilde\phi_+\big(\eta+a_\Gamma(\mu_s,\mu),\mu_s\big) \,.
\end{align}
On the right-hand side of this relation the Laplace-transformed LCDA is evaluated at a shifted value of $\eta$. While the moments $\lambda_B$ and $\sigma_n$ needed in fixed-order perturbation theory are given in terms of the Laplace transform and its derivative at the origin, see (\ref{origins}), taking into account resummation effects requires knowledge of the function $\tilde\phi_+(\eta,\mu)$ away from the origin. Indeed, the evolution equation (\ref{wonderful}) shows that performing a scale transformation shifts the value of $\eta$. 

The positions of the nearest singularities at positive (negative) values of $\eta$ determine the asymptotic behavior of the momentum-space LCDA for small (large) values of $\omega$ \cite{Lange:2003ff}. At the low scale $\mu_s$ we denote these values by $\eta_+$ and $-\eta_-$. When the LCDA is evolved to a higher scale, the positions of these singularities shift to $\eta_+ +|a_\Gamma(\mu_s,\mu)|$ and $-\eta_- +|a_\Gamma(\mu_s,\mu)|$, taking into account that $a_\Gamma(\mu_s,\mu)<0$ for $\mu>\mu_s$. Additional singularities are generated by the $\Gamma$-functions in the numerator of (\ref{greatsolu}) and are located at $\eta=1$ and $\eta=-1+|a_\Gamma(\mu_s,\mu)|$. For sufficiently large values of $\mu$ the nearest positive singularity is the one at $\eta=1$, corresponding to a linear behavior $\phi_+^B(\omega,\mu)\sim\omega$ near the origin. The nearest negative singularity is located at $\eta=-\min(1,\eta_-)+|a_\Gamma(\mu_s,\mu)|$, implying that $\phi_+^B(\omega,\mu)$ falls off slower than $1/\omega$ at large~$\omega$.

The integral in the exponent of the term in the second line of (\ref{greatsolu}) can be expanded in powers of $\alpha_s$, because the function $\G$ starts at ${\cal O}(\alpha_s^2)$. In the beautiful papers \cite{Braun:2014owa,Braun:2018fiz} it was shown that the Lange-Neubert kernel for the $B$-meson LCDA in (\ref{gammaplus}) can be written in a remarkably compact form as a logarithm of the generator of special conformal transformations along the light-cone. Using tools from conformal field theory the transformation of the evolution equation (\ref{RGEphiB}) to the so-called ``dual space'', originally proposed in \cite{Bell:2013tfa}, was rederived. In subsequent work the evolution equation was extended to two-loop order \cite{Braun:2016qlg,Braun:2019wyx}. After the conversion back to momentum space, one finds \cite{Liu:2020ydl}
\begin{align}\label{2loopgammas}
   \gamma(\alpha_s)
   &= - \frac{C_F\alpha_s}{2\pi} 
    - \left( 5.523 - 0.358\hspace{0.5mm}n_f \right) \left( \frac{\alpha_s}{\pi} \right)^2 , \notag \\
   \hat\gamma_+(\omega,\omega';\alpha_s)
   &= C_F \left( \frac{\alpha_s}{2\pi} \right)^2 
    \frac{\omega\,\theta(\omega'-\omega)}{\omega'(\omega'-\omega)}\,
    h\Big(\frac{\omega}{\omega'}\Big) \,,
\end{align}
where the numerical value of the two-loop coefficient of $\gamma$ corresponds to $N_c=3$ colors, and
\begin{equation}
   h(x) = \ln x \left[ \beta_0 
    + 2C_F\!\left( \ln x - \frac{1+x}{x}\,\ln(1-x) - \frac32 \right) \right] .
\end{equation}
It is then straightforward to express the two-loop contribution to $\G(\eta,\alpha_s)$ in terms of the digamma function $\psi(1-\eta)$ and its derivative. For the exponent in the second line of (\ref{greatsolu}) we find, with $r=\alpha_s(\mu_s)/\alpha_s(\mu)$,
\begin{equation}
   \frac{C_F\alpha_s(\mu)}{2\pi}\,\int_0^1\!\frac{dx}{1-x}\,\frac{h(x)}{\beta_0}\,x^{-\eta}\,
    \frac{r^{1+\frac{2C_F}{\beta_0} \ln x}-1}{1+\frac{2C_F}{\beta_0} \ln x} + {\cal O}(\alpha_s^2) \,.
\end{equation}

Expanding the evolution equation (\ref{wonderful}) about $\eta=0$ one can derive a coupled, infinite set of evolution equations for $\lambda_B$ and the logarithmic moments $\sigma_n$ defined in (\ref{moments}). The first few relations are 
\begin{align}
   \frac{d\ln\!\lambda_B(\mu)}{d\ln\mu}
   &= \Gamma_c(\alpha_s) \left[ \ln\frac{\mu}{\bar\omega} + \sigma_1(\mu) \right] 
    + \gamma(\alpha_s) - \G(0,\alpha_s) \,, \notag \\
   \frac{d\sigma_1(\mu)}{d\ln\mu}
   &= \Gamma_c(\alpha_s) \left[ \sigma_1^2(\mu) - \sigma_2(\mu) \right] 
    +  \G^{(1)}(0,\alpha_s) \,, \notag \\[1mm]
   \frac{d\sigma_2(\mu)}{d\ln\mu}
   &= \Gamma_c(\alpha_s) \left[ \sigma_1(\mu)\,\sigma_2(\mu) - \sigma_3(\mu) + 4\zeta_3 \right] 
    \notag \\
   &\quad\mbox{}+ 2\sigma_1(\mu)\,\G^{(1)}(0,\alpha_s) + \G^{(2)}(0,\alpha_s) \,.
\end{align}
They are derived here for the first time. The fact that this system does not close hints at the fact that to control the evolution of the moments one needs to know the behavior of the Laplace transform in the vicinity of the origin, which is equivalent to an infinite set of moments. The exact solution to the above equations can be obtained from the expansion of (\ref{greatsolu}) about $\eta=0$. In particular, we find 
\begin{align}
   \lambda_B^{-1}(\mu)
   &= N(\mu_s,\mu)\,e^{2\gamma_E a_\Gamma(\mu_s,\mu)}\,
    \frac{\Gamma\big(1+a_\Gamma(\mu_s,\mu)\big)}{\Gamma\big(1-a_\Gamma(\mu_s,\mu)\big)} \\[-1mm]
   &\hspace{-0.9cm}\times
    \exp\Bigg[\,\int\limits_{\alpha_s(\mu_s)}^{\alpha_s(\mu)}\!\frac{d\alpha}{\beta(\alpha)}\,
    \G\big(a_\Gamma(\mu_\alpha,\mu),\alpha\big) \Bigg]\, 
    \tilde\phi_+\big(a_\Gamma(\mu_s,\mu),\mu_s\big) \,. \notag
\end{align}

\section{Unbiased Parameterization of the LCDA} 

For practical purposes one needs a parameterization of the Laplace transform $\tilde\phi_+(\eta,\mu_s)$ at the low scale $\mu_s$, which ideally should be free of unjustified assumptions. Without loss of generality, we choose the parameter $\bar\omega$ in (\ref{Laplace}) such that the first moment vanishes at this scale, $\sigma_1(\mu_s)=0$. In essence, we trade the hadronic parameter $\sigma_1$ for a new parameter $\bar\omega\sim\Lambda_{\rm QCD}$. According to (\ref{moments}) this defines $\bar\omega$ via the average value of the distribution amplitude $\phi_+^B(\omega,\mu_s)$ in the variable $\ln\omega$. With this choice it is likely that the higher moments do not take unnaturally large values either. We thus obtain the unbiased parameterization 
\begin{equation}\label{series}
   \tilde\phi_+(\eta,\mu_s)
   = \frac{1}{\lambda_B(\mu_s)}\,\bigg[ 1 + \sum_{n\ge 2}\,\frac{\eta^n}{n!}\,\sigma_n(\mu_s)
    \bigg] \,,
\end{equation}
in which the parameters $\lambda_B$, $\bar\omega$ and $\sigma_n$ are uncorrelated. If the required region of $\eta$ values is such that $|\eta|\ll 1$, then the first few terms in this series should be sufficient to obtain a reliable approximation. 

\begin{figure}
\begin{center}
\vspace{2mm}
\includegraphics[width=0.45\textwidth]{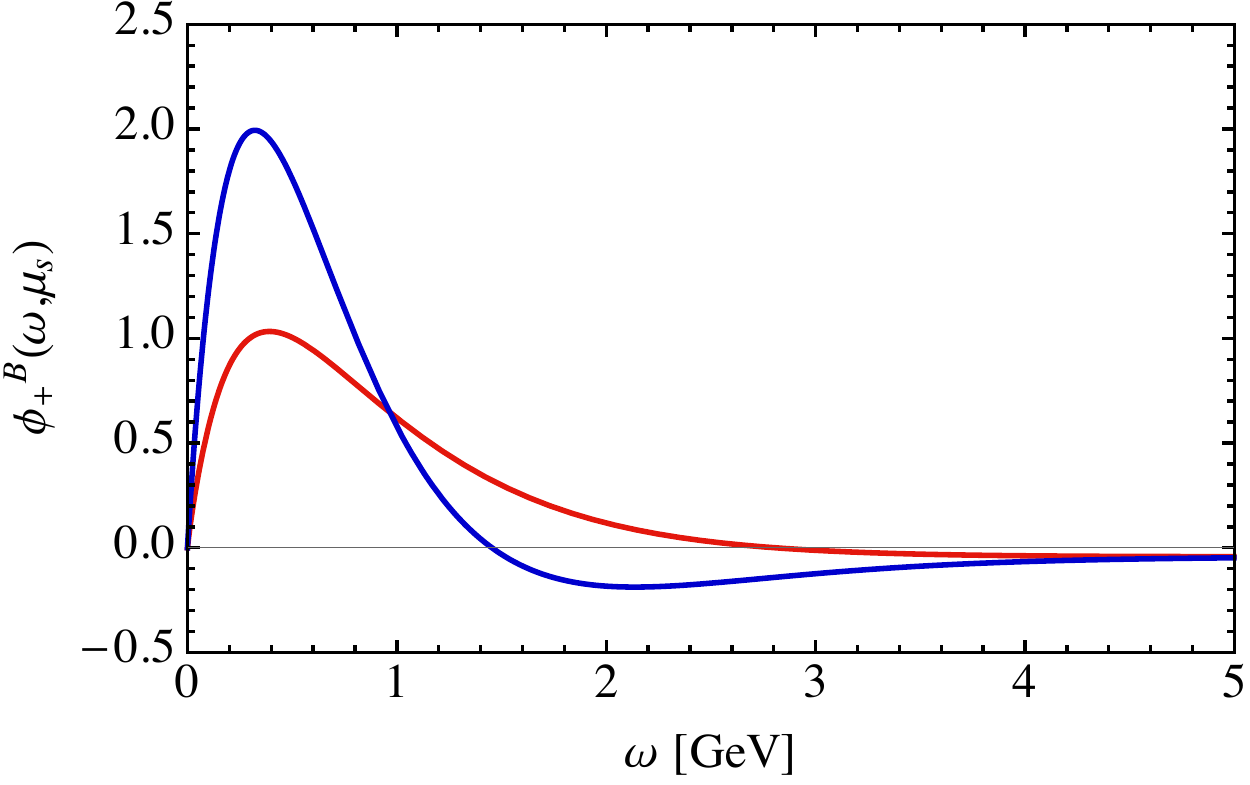}
\includegraphics[width=0.45\textwidth]{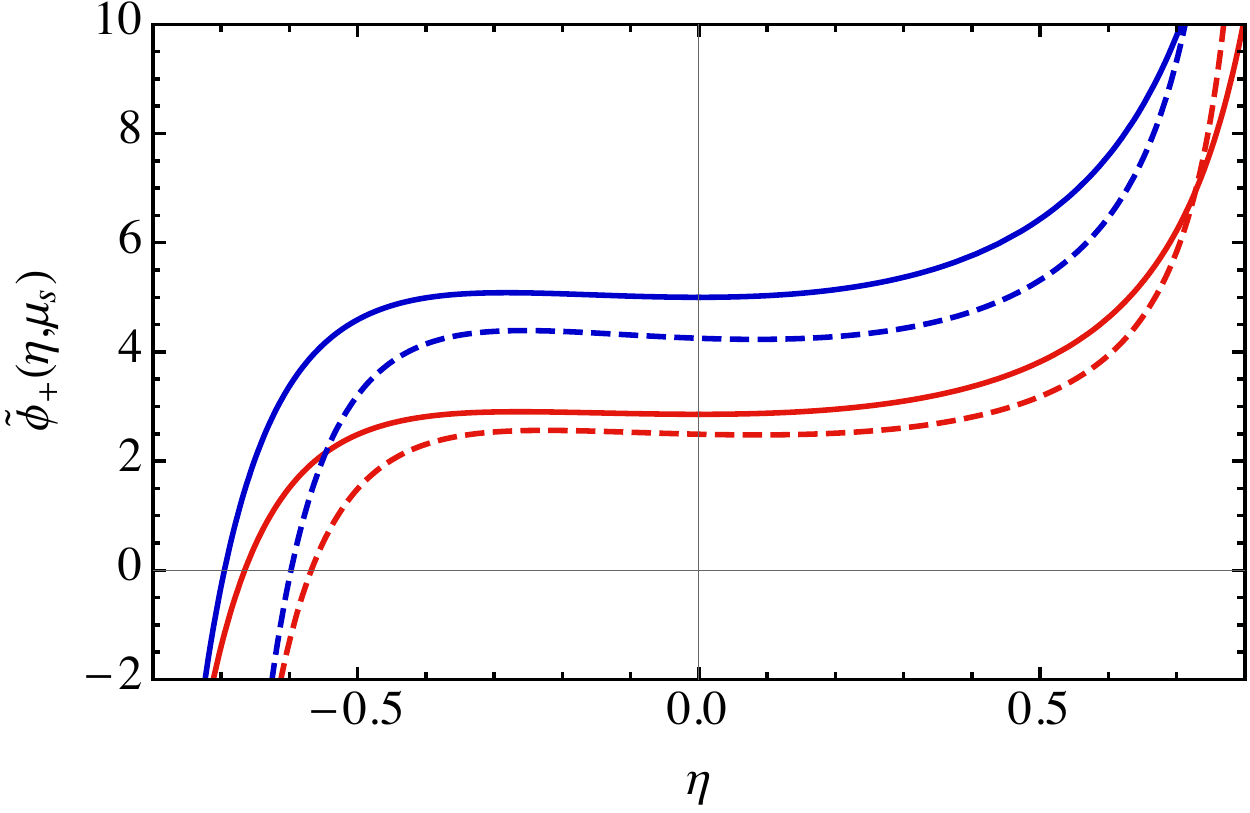}
\end{center}
\vspace{-0.5cm}
\caption{\label{fig:LCDA} 
Model functions for the LCDA at the scale $\mu_s=1$\,GeV in momentum space (top) and in Laplace space (bottom). The curves differ in the choice of $b=0$ (red) and $b=-2.07$ (blue). The dashed lines in the lower plot show the results at the higher scale $\mu=1.5$\,GeV.}
\end{figure}

It is instructive to illustrate our results with a numerical example. At the low scale $\mu_s=1$\,GeV we impose the two-parameter model function 
\begin{equation}
\begin{aligned}
   \phi_+^B(\omega,\mu_s) 
   &= \left( 1 - b + \frac{b\omega}{2\omega_0} \right) \frac{\omega}{\omega_0^2}\,
    e^{-\omega/\omega_0} \\
   &\quad\mbox{}+ \frac{4\alpha_s(\mu_s)}{3\pi}\,\frac{\omega}{\omega^2+\omega_0^2} 
    \left( \frac12 - \ln\frac{\omega}{\mu_s} \right) ,
\end{aligned}
\end{equation}
which satisfies most known properties of the LCDA. The first component on the right-hand side exhibits an  exponential fall-off and is normalized to~1. The second component correctly takes into account the radiative tail for large $\omega$ values \cite{Lee:2005gza}, which renders the integral over the LCDA divergent. (For simplicity we ignore power-suppressed contributions to the tail from higher-dimensional operators \cite{Lee:2005gza,Kawamura:2008vq}.) We take $\omega_0=482$\,MeV and consider the two choices $b=0$ and $b=-2.07$, for which the first integral in (\ref{moments}) yields $\lambda_B(\mu_s)\simeq 350$\,MeV and 200\,MeV, respectively. While a value around 350\,MeV is often considered as a default choice for $\lambda_B$, in phenomenological applications of the QCD factorization approach to non-leptonic $B$ decays one typically prefers a lower value around 200\,MeV (see e.g.\ scenarios S2 and S4 in \cite{Beneke:2003zv}). It is straightforward to calculate the Laplace transforms $\tilde\phi_+(\eta,\mu_s)$ of the two model functions and the associated parameters $\bar\omega$, for which we obtain $\bar\omega\simeq 183$\,MeV and 141\,MeV. The first few moments are $\sigma_2(\mu_s)\simeq 1.17$, $\sigma_3(\mu_s)\simeq 6.41$ and $\sigma_4(\mu_s)\simeq -6.88$ for $b=0$, and $\sigma_2(\mu_s)\simeq 1.04$, $\sigma_3(\mu_s)\simeq 5.32$ and $\sigma_4(\mu_s)\simeq -3.90$ for $b=-2.07$. Note that $\sigma_4$ is negative in both cases. The top panel in Figure~\ref{fig:LCDA} shows the two model functions in momentum space, while the bottom panel shows the corresponding Laplace transforms. They exhibit pole singularities at $\eta=\pm 1$. While the two functions look rather different in momentum space, their Laplace images are very similar apart from a shift in the vertical direction, reflecting the two different values of $\lambda_B$. In other words, allowing for negative $b$ values is a very effective means to lowering $\lambda_B$ while keeping the moments $\sigma_n$ approximately unaffected. Note that, owing to our choice $\sigma_1(\mu_s)=0$, the Laplace transforms exhibit a flat behavior for $|\eta|\lesssim 0.3$, so that in this region they can be well approximated by keeping the first few terms in the series expansion (\ref{series}). The dashed lines in the lower plot show the RG-evolved functions $\tilde\phi(\eta,\mu)$ at the higher scale $\mu=1.5$\,GeV, obtained from (\ref{greatsolu}). The two main effects of RG evolution are the modest increase of $\lambda_B(\mu)$ and the shift of the nearest singularity at negative $\eta$ from $-1$ to $-1+|a_\Gamma(\mu_s,\mu)|\simeq-0.9$, in accordance with our discussion following relation (\ref{greatsolu}). Note that in our results below we only use the model-independent parameterization (\ref{series}) and make no reference to the particular model considered above.

\section{Scale-Independent Factorization Formula} 

With the help of the exact solution (\ref{greatsolu}) and the known solutions of the evolution equations for the hard function and jet function \cite{Liu:2020ydl}, we have derived an all-order formula for the convolution integral in (\ref{fact}), in which any reference to the factorization scale $\mu$ cancels out explicitly and in which all large logarithms are resummed. We find
\begin{widetext}
\begin{equation}\label{master}
\begin{aligned}
   I = \int_0^\infty\!\frac{d\omega}{\omega}\,T(m_b,E_\gamma,\omega,\mu)\,\phi_+^B(\omega,\mu)
   &= \exp\Big[ S(\mu_h,\mu_j) + S(\mu_s,\mu_j) - a_{\gamma_H}(\mu_h,\mu_j) 
    + a_\gamma(\mu_s,\mu_j) + 2\gamma_E\hspace{0.3mm}a_\Gamma(\mu_s,\mu_j) \Big] \\[1mm]
   &\hspace{-3.85cm}\times
    H(m_b,E_\gamma,\mu_h) \left( \frac{2E_\gamma}{\mu_h} \right)^{-a_\Gamma(\mu_h,\mu_j)}
    \J(\partial_\eta,\mu_j)\,\bigg( \frac{2E_\gamma\hspace{0.2mm}\bar\omega}{\mu_j^2} \bigg)^\eta\,\,
    \frac{\Gamma\big(1-\eta+a_\Gamma(\mu_s,\mu_j)\big)\,\Gamma(1+\eta)}%
         {\Gamma\big(1+\eta-a_\Gamma(\mu_s,\mu_j)\big)\,\Gamma(1-\eta)} \\[-1mm]
   &\hspace{-3.85cm}\times
    \exp\Bigg[\,\int\limits_{\alpha_s(\mu_s)}^{\alpha_s(\mu_j)}\!
    \frac{d\alpha}{\beta(\alpha)}\,\G\big(-\eta+a_\Gamma(\mu_\alpha,\mu_j),\alpha\big) \Bigg] 
    \left( \frac{\bar\omega}{\mu_s} \right)^{-a_\Gamma(\mu_s,\mu_j)} 
    \tilde\phi_+\big(\!-\!\eta+a_\Gamma(\mu_s,\mu_j),\mu_s\big)\, \bigg|_{\eta=0} \,.
\end{aligned}
\end{equation}
\end{widetext}
This elegant formula is vastly simpler than the corresponding relation in momentum space, which is currently only known at NLO \cite{Bosch:2003fc,Beneke:2011nf}. Its extension to NNLO would involve a three-dimensional integral over an integrand featuring a very complicated dependence on $\mu$ \cite{Liu:2020eqe}, despite the fact that the result obtained after performing the integrations numerically is formally $\mu$ independent. The master formula (\ref{master}) involves the matching conditions for the hard function $H$ at a scale $\mu_h\sim m_b$ and for the jet function $\J(L_p,\mu_j)\equiv J(-p^2,\mu_j)$ with $L_p=\ln(p^2/\mu_j^2)$ at a scale $\mu_j\sim\sqrt{m_b\Lambda_{\rm QCD}}$, which must be provided by perturbation theory. Both are known at two-loop order. The hard function $H=C_1/K_F$ is the ratio of the matching coefficient $C_1$ of the heavy-light current operator \cite{Bonciani:2008wf,Asatrian:2008uk,Beneke:2008ei,Bell:2008ws} and the matching coefficient $K_F$ of the $B$-meson decay constant \cite{Broadhurst:1994se,Grozin:1998kf}. The two-loop expression for the jet function was recently derived in \cite{Liu:2020ydl}. In (\ref{master}) the first argument of $\J(L_p,\mu_j)$ is replace by a derivative operator $\partial_\eta$ with respect to an auxiliary parameter $\eta$, which acts on all terms standing to the right. The solution also contains the initial condition for the Laplace transform of the LCDA at the low scale $\mu_s$. With a typical choice $\mu_s=1$\,GeV and $\mu_j=1.5$\,GeV one finds $a_\Gamma(\mu_s,\mu_j)\simeq -0.098$, confirming our claim that in the solution one needs the Laplace transform in the region close to the origin. Note that the master formula (\ref{master}) is formally independent of the matching scales $\mu_h$, $\mu_j$ and $\mu_s$. The RG functions $S(\mu_1,\mu_2)$, $a_\Gamma(\mu_1,\mu_2)$ and $a_{\gamma_H}(\mu_1,\mu_2)$ are computed at NNLO using the four-loop cusp anomalous dimension \cite{Henn:2019swt} and the three-loop anomalous dimension of the hard function derived from \cite{Chetyrkin:2003vi,Becher:2009qa,Bruser:2019yjk}. The functions $a_\gamma(\mu_1,\mu_2)$ and the integral over $\G$ are evaluated using the known two-loop anomalous dimensions $\gamma$ and $\hat\gamma_+$ \cite{Braun:2019wyx}, which is sufficient since the two scales $\mu_s$ and $\mu_j$ lie rather close to each other. We are thus in a position to evaluate the convolution integral at NNLO in RG-improved perturbation theory. 

\begin{figure}
\begin{center}
\vspace{1.3mm}
\includegraphics[width=0.45\textwidth]{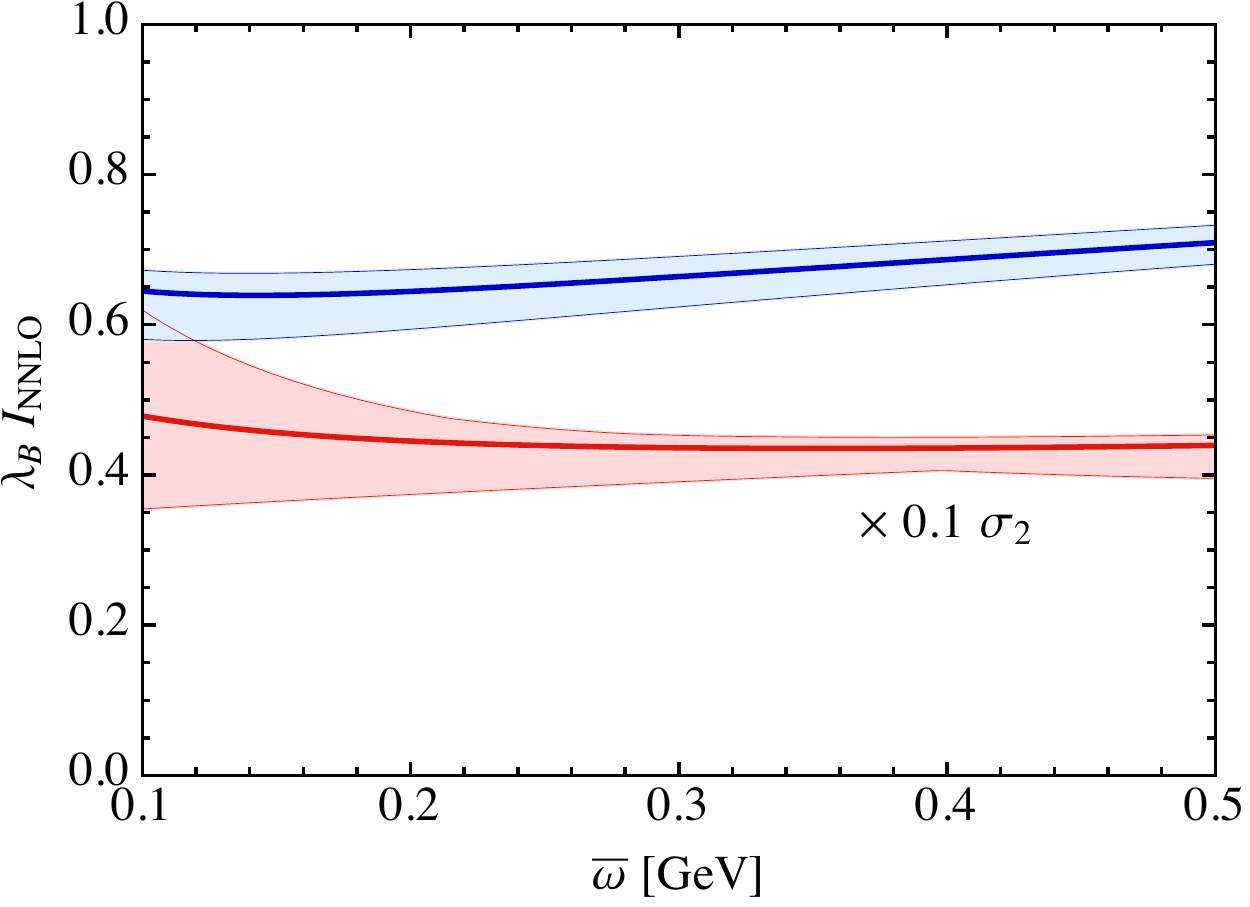}
\end{center}
\vspace{-5mm}
\caption{\label{fig:wbar} 
Coefficients of the leading term (blue) and of $0.1\sigma_2$ (red) in the result (\ref{numbers}) including scale variations added in quadrature, for different values of $\bar\omega$.}
\end{figure}

To illustrate our results, we fix the photon energy at $E_\gamma=2.2$\,GeV, which is at the center of the region between 2 and 2.4\,GeV, where the factorization theorem (\ref{fact}) can be applied safely. The energy dependence of our numerical results in this region is very weak. Following \cite{Beneke:2018wjp} we define the hadronic input parameters $\lambda_B$, $\bar\omega$ and $\sigma_n$ at the low scale $\mu_s=1$\,GeV, vary the hard matching scale by a factor~2 about the default value $\mu_h=m_b$, and vary the intermediate matching scale by a factor $\sqrt{2}$ about the default value $\mu_j=\sqrt{2}$\,GeV. The smaller scale variation for $\mu_j$ is justified due to the fact that parametrically $\mu_j^2\sim\mu_h\hspace{0.3mm}\mu_s$ and we keep $\mu_s$ fixed at 1\,GeV. Working consistently at NNLO in RG-improved perturbation theory, we obtain
\begin{equation}\label{numbers}
\begin{aligned}
   I_{\rm NNLO} = \frac{1}{\lambda_B}\, 
   & \Big[ \big( 0.664\,_{-0.013}^{+0.011}\,_{-0.038}^{+0.024} \big) \\
   &\hspace{1mm}+ \big( 4.36\,_{-0.08}^{+0.15}\,_{-0.45}^{+0.07} \big)
    \cdot 10^{-2}\,\sigma_2 \\[1mm]
   &\hspace{1mm}+ \big( 0.35\,_{-0.02}^{+0.12}\,_{-1.99}^{+2.97} \big)
    \cdot 10^{-3}\,\sigma_3 \\
   &\hspace{1mm}+ \big( 5.02\,_{-0.08}^{+0.28}\,_{-1.91}^{+5.84} \big)
    \cdot 10^{-4}\,\sigma_4 + \dots \Big] \,,
\end{aligned}
\end{equation}
where for each value the quoted errors arise from the variations of $\mu_h$ and $\mu_j$. Our central value 0.664 is about 10\% smaller than the central value of the NLO result
\begin{equation}
   I_{\rm NLO} = \frac{1}{\lambda_B}\,\big[ 0.731 + 0.035\,\sigma_2 
    - 0.003\,\sigma_3 + \dots \big] \,.
\end{equation}
The NNLO corrections included here for the first time thus have a significant impact on the $B^-\to\gamma\ell^-\bar\nu$ branching ratio. The result (\ref{numbers}) refers to $\bar\omega=300$\,MeV. Figure~\ref{fig:wbar} shows how the coefficients of the leading term and of $\sigma_2$ vary with $\bar\omega$. The range shown is motivated by the fact that parametrically $\bar\omega\sim\Lambda_{\rm QCD}$. The leading coefficient increases slightly with $\bar\omega$, whereas the coefficient of $\sigma_2$ is almost independent of it. Note that the scale variations increase for smaller values of $\bar\omega$. As can be seen from (\ref{greatsolu}), the quantity $2E_\gamma\bar\omega$ sets the ``natural'' scale for $\mu_j^2$, and for $\bar\omega<0.23$\,GeV this scale drops below 1\,GeV, outside the range of variation of $\mu_j$. This suggests that the perturbative corrections to the jet function get larger the smaller $\bar\omega$ is.

\section{Conclusions} 

In summary, we have shown that the information about the $B$-meson LCDA that can be probed in hard exclusive processes such as $B^-\to\gamma\ell^-\bar\nu$ is entirely and most directly contained in the Laplace transform $\tilde\phi_+(\eta,\mu)$. We have obtained the RG evolution equation satisfied by this function and presented its exact solution. Using this result, we have derived a closed analytic expression for the RG-improved form of the convolution integral governing the $B^-\to\gamma\ell^-\bar\nu$ decay amplitude at leading power in $\Lambda_{\rm QCD}/m_b$. Finally, we have proposed an unbiased parameterization of $\tilde\phi_+(\eta)$ in terms of uncorrelated hadronic parameters $\lambda_B$, $\bar\omega$ and $\sigma_{n\ge 2}$ defined at the scale $\mu_s$. Our results provide the basis for an accurate determination of the important parameter $\lambda_B$ from future high-precision measurements of the $B^-\to\gamma\ell^-\bar\nu$ photon energy spectrum in the region near the kinematic endpoint. To this end, it will however be important to also include power corrections in $\Lambda_{\rm QCD}/m_b$, a detailed study of which has been presented in \cite{Beneke:2011nf,Braun:2012kp,Wang:2016qii,Beneke:2018wjp}.

\vspace{2mm}
{\em Acknowledgements:\/} We are grateful to Andrey Grozin for useful discussions concerning the two-loop matching coefficient $K_F$ and to Ben Pecjak for providing us with a MATHEMATICA implementation of the two-loop matching coefficient for the heavy-light current. This work has been supported by the Cluster of Excellence {\em Precision Physics, Fundamental Interactions, and Structure of Matter} (PRISMA$^+$\! EXC 2118/1) funded by the German Research Foundation (DFG) within the German Excellence Strategy (Project ID 39083149).

\end{document}